\documentstyle[12pt,cite,epsf, amssym]{article}
\renewcommand{\thefootnote}{\fnsymbol{footnote}}
\def \be{\begin{equation}}
\def \ee{\end{equation}}
\def \bea{\begin{eqnarray}}
\def \eea{\end{eqnarray}}
\def \ben{\begin{enumerate}}
\def \een{\end{enumerate}}
\def \asym{A_{\mathrm {CP}}}
\def \bem#1{\renewcommand{\thefootnote}{\arabic{footnote}}\footnote{#1}}
\def \bm{\boldmath}

\def \branch{\mathrm {Br}\,}

\def \cp{{\it CP\/}}
\def \cpbf{ \bm$C$\hspace{-0.1em}\bm$P$\/}

\def \ea{{\it et al.\/}}

\def \eq#1{Eq.~(\ref{#1})}
\def \eqs#1#2{Eqs.~(\ref{#1})--(\ref{#2})}

\def \GeV{{\mathrm{GeV}}}

\def \heff{H_{\mathrm{eff}}}
\def \Im{{\mathrm{Im}}\,}

\def \MeV{{\mathrm{MeV}}}

\def \nnu{\nonumber}

\def \O{{\cal O}}
\def \ol#1{\overline{#1}}

\def \rcontcc{R_{\mathrm {cont}}^{c\bar{c}}(\sh)}
\def \rcontuu{R_{\mathrm {cont}}^{u\bar{u}+d\bar{d}}(\sh)}

\def \rhadronrho{R_{\mathrm {had}}^{\r}(\sh)}
\def \rhadronomega{R_{\mathrm {had}}^{\o}(\sh)}
\def \rhadronpsi{R_{\mathrm {had}}^{J/\psi}(\sh)}
\def \rhadron{R_{\mathrm {had}}}

\def \rresrho{R_{\mathrm {res}}^{\r}(\sh)}
\def \rresomega{R_{\mathrm {res}}^{\o}(\sh)}
\def \rrespsi{R_{\mathrm {res}}^{J/\psi}(\sh)}
\def \Re{{\mathrm{Re}}\,}
\def \rf{Ref.~\cite}
\def \rfs{Refs.~\cite}

\newcommand{\TO}[2]{\stackrel {#1}{\hbox to #2pt{\rightarrowfill}}}

\def \cseff {c_7^{\mathrm{eff}}}
\def \ceff {c_9^{\mathrm{eff}}}
\def \c9eff*{c_9^{\mathrm{eff}*}}
\def \a{\alpha}
\def \b{\beta}
\def \g{\gamma}
\def \G{\Gamma}

\def \k{\kappa}
\def \l{\lambda}

\def \m{\mu}
\def \n{\nu}
\def \o{\omega}
\def \p{\pi}
\def \r{\rho}
\def \s{\sigma}

\def \t{\tau}
\def \mbh{\hat{m}_b}
\def \mch{\hat{m}_c}

\def \mch{\hat{m}_c}
\def \mdh{\hat{m}_d}
\def \mh{\hat{m}}
\def \mlh{\hat{m}_l}

\def \muh{\hat{m}_u}

\def \sh{\hat{s}}

\def \today{\ifcase\month\or
  January\or February\or March\or April\or May\or June\or
  July\or August\or September\or October\or November\or December\fi
  \space\number\year}
\newcounter{Section}
\def \theSection{\Roman{Section}} 
\newcommand{\Sec}[1]{\refstepcounter{Section}%
\centerline{\bf\theSection. #1}\setcounter{equation}{0}%
\renewcommand{\theequation}{\arabic{Section}.\arabic{equation}}}
\newcounter{Subsection}[Section]

\newcounter{Subsubsection}[Section]

\newcounter{Appendix}
\def \theAppendix{\Alph{Appendix}}
\newcommand{\app}[1]{\refstepcounter{Appendix}%
\centerline{\bf APPENDIX \theAppendix: #1}\setcounter{equation}{0}%
\renewcommand{\theequation}{\Alph{Appendix}\arabic{equation}}}

\def\ib#1#2#3{{\it ibid.\/}~{\bf#1} (19#2) #3}

\def\np#1#2#3{{\it Nucl.~Phys.\/}~{\bf B#1} (19#2) #3}
\def\pl#1#2#3{{\it Phys.~Lett.\/}~{\bf B#1} (19#2) #3}
\def\pp{{\it preprint\/} }
\def\prd#1#2#3{{\it Phys.~Rev.\/}~{\bf D#1} (19#2) #3}
\def\prl#1#2#3{{\it Phys.~Rev.~Lett.\/}~{\bf #1} (19#2) #3}

\def\zpc#1#2#3{{\it Z.~Phys.\/}~{\bf C#1} (19#2) #3}
\begin{document}
\begin{flushright}
PITHA 96/25 \\ hep-ph/9608361 \\ August 1996
\end{flushright}
\begin{center}
\LARGE \bf \cpbf\ Violation  in the Decay \bm{$B\to X_d \,e^+ e^-$}
\end{center}
\begin{center}
\sc F.~Kr\"uger\footnote{\footnotesize
Electronic address: krueger@physik.rwth-aachen.de} and
L.\,M.~Sehgal\footnote{\footnotesize Electronic address: sehgal@physik.rwth-aachen.de}
\\ \it Institut f\"ur Theoretische Physik (E), RWTH Aachen\\
D-52056 Aachen, Germany
\end{center}
\vspace{1.0cm}
\thispagestyle{empty}
\centerline{\bf ABSTRACT}
\begin{quotation}
The decay $b\to d\,e^+e^-$ has an amplitude containing comparable contributions proportional to 
$V_{tb}^{}V_{td}^{*}$, $V_{cb}^{}V_{cd}^{*}$ and $V_{ub}^{}V_{ud}^{*}$. These pieces involve
different unitarity phases produced by $c\bar{c}$ and $u\bar{u}$ loops. The simultaneous presence of different CKM phases and different dynamical phases leads to a calculable asymmetry in the partial widths
of $b\to d\,e^+e^-$ and $\bar{b}\to \bar{d}\,e^+e^-$. Using the effective Hamiltonian of the standard model, we calculate this asymmetry as a function of the $e^+e^-$ invariant mass. The effects of $\r$, $\o$ and $J/\psi$
resonances are taken into account in the vacuum polarization of the $u\bar{u}$ and $c\bar{c}$ currents. As a typical result, an asymmetry of $- 5\%$ ($- 2\%$) is predicted in the nonresonant domain 
$1\ \GeV < m_{e^+e^-} < m_{J/\psi}$, assuming $\eta = 0.34$ and $\r= 0.3$ ($-0.3$). 
The branching ratio in this domain is $1.2\times 10^{-7}$ ($3.3\times 10^{-7}$). Results are also obtained in the region of the $J/\psi$ resonance, where an asymmetry of $3\times 10^{-3}$ is expected, subject to certain theoretical uncertainties in the $b\to d J/\psi$ amplitude.
\end{quotation}
\begin{quote}
\end{quote}
\setcounter{footnote}{0}
%
%
\newpage
\Sec{INTRODUCTION} \label{intro}
The decays $B\to X_{s, d}\, l^+l^-$ are important probes of the effective Hamiltonian governing the 
flavour-changing neutral current transition $b\to s (d)\, l^+l^-$ \cite{alirev}. 
The matrix element contains a term describing the virtual effects of the top quark proportional to 
 $V_{tb}^{}V_{tq}^{*}$, $q = s, d$, and in addition terms induced by $c\bar{c}$ and 
$u\bar{u}$ loops, proportional to $V_{cb}^{}V_{cq}^{*}$ and $V_{ub}^{}V_{uq}^{*}$. In the case of the 
decay $b\to s\, l^+l^-$, the relevant CKM factors have the order of magnitude 
$V_{tb}^{}V_{ts}^{*}\sim \l^3$, $V_{cb}^{}V_{cs}^{*}\sim \l^3$, $V_{ub}^{}V_{us}^{*}\sim \l^5$, where
$\l = \sin\theta_{\mathrm C}\simeq 0.221$. This has the consequence that the $u\bar{u}$ contribution is very small, and the unitarity relation for the CKM factors reduces approximately to 
$V_{tb}^{}V_{ts}^{*}+V_{cb}^{}V_{cs}^{*} \approx 0$. Thus the effective Hamiltonian for $b\to s\, l^+l^-$ essentially involves only one independent CKM factor $V_{tb}^{}V_{ts}^{*}$, so that \cp\ violation 
in this channel is strongly suppressed, within the standard model \cite{aliev, duyang}.

The situation is quite different for the transition $b\to d\, l^+l^-$. The internal top-quark contribution is proportional to  $V_{tb}^{}V_{td}^{*}$,  while the  terms related to $c\bar{c}$
and $u\bar{u}$ loops are proportional to $V_{cb}^{}V_{cd}^{*}$ and $V_{ub}^{}V_{ud}^{*}$.
All of these CKM factors are of order $\l^4$, and, a priori, can have quite different phases. In addition, the 
$c\bar{c}$ and $u\bar{u}$ loop contributions are accompanied by different unitarity phases corresponding 
to real intermediate states. We thus have a situation in which the amplitude contains pieces with different
CKM phases as well as different dynamical (unitarity) phases. These are precisely the  desiderata for observing 
\cp-violating asymmetries in partial rates. The purpose of this paper is to derive quantitative predictions for the \cp-violating partial
width asymmetry between the channels $b\to d\, e^+e^-$ and $\bar{b}\to \bar{d}\, e^+e^-$.
\vspace{1.5cm}

\Sec{THE EFFECTIVE HAMILTONIAN FOR \bm $b\to d \,l^+ l^-$}
The effective Hamiltonian for the decay $b\to d \,l^+ l^-$ in the standard model can be written as 
\bea\label{heff}
\heff &=& -\frac{4G_F}{\sqrt{2}}V_{tb}^{}V_{td}^{*}\, \Bigg\{\sum\limits_{i = 1}^{10} 
c_i (\m)\O_i(\m)\nnu\\ 
& &-\l_u\Big [c_1(\m)  \Big(\O_1^u(\m) - \O_1(\m)\Big) + c_2(\m)  \Big(\O_2^u(\m) - \O_2(\m)\Big)
\Big ]\Bigg\}\ ,
\eea
where we have used the unitarity of the Cabibbo-Kobayashi-Maskawa (CKM) matrix  
$V_{tb}^{}V_{td}^{*}+V_{ub}^{}V_{ud}^{*}=-V_{cb}^{}V_{cd}^{*}$, 
and $\l_u \equiv V_{ub}^{}V_{ud}^{*}/V_{tb}^{}V_{td}^{*}$.
For the purpose of this paper it is convenient  to use the Wolfenstein representation \cite{wolf} of the 
CKM matrix with  four real parameters $\l =  \sin\theta_{\mathrm C}\simeq 0.221$, $A$, $\rho$, 
and $\eta$, where $\eta$ is a measure of \cp\ violation. In terms of these parameters  
\bea\label{ckmwolf}
\l_u =\frac{\rho(1-\rho)-\eta^2}{(1-\rho)^2 + \eta^2} - i \frac{\eta}{(1-\rho)^2 + \eta^2} + \cdots\ , 
\eea
where the ellipsis denotes higher-order terms in $\l$. Furthermore, we will make use of
\bea\label{ckmwolf1}
\frac{| V_{tb}^{}V_{td}^{*}|^2}{|V_{cb}|^2} = \l^2 \left( (1-\rho)^2 + \eta^2\right) + O(\l^4)\ .
\eea
The operator basis $\{\O_i \}$ for $\heff$ is given in \rfs{gsw, misiak} with the obvious replacement 
$s\to d$, and the additional operators $\O_{1, 2}^u$ read
\bea
\O_1^u = (\bar{d}_{\a}\g_{\m} P_Lu_{\b}) (\bar{u}_{\b}\g^{\m} P_L b_{\a}), 
\quad \O_2^u = (\bar{d}_{\a}\g_{\m} P_Lu_{\a}) (\bar{u}_{\b}\g^{\m} P_L b_{\b})\ ,
\eea 
with $P_{L,R} =(1\mp\g_5)/2$.
The evolution of the Wilson coefficients $c_i(\m)$ in \eq{heff} from the scale $\m = m_W$ 
down to $\m= m_b$ by means of the renormalization group equation has been discussed in several papers, 
and we refer the reader to the review article of Buchalla \ea \cite{burev}. 
The resulting QCD-corrected matrix element  can be written as
\bea\label{matrixele}
\cal{M}&=&\left.\frac{4 G_F}{\sqrt{2}}V_{tb}^{}V_{td}^*\,\frac{\a}{4\p}\right\{\ceff ( \bar{d}\g_{\m}
P_L b )
\bar{l}\g^{\m}l+ c_{10}( \bar{d}\g_{\m}P_L b )\bar{l}\g^{\m}\g^5 l\nnu\\
& &\mbox{}-2\,\cseff \bar{d}\left. i  \sigma_{\m\n}\frac{q^{\n}}{q^2}\left(m_b P_R + m_d P_L\right)b\, 
\bar{l}\g^{\m} l\right\} .
\eea
Neglecting terms of 
$O(m_q^2/m_W^2)$, $q = u, d, c$, the analytic expressions for all Wilson coefficients, except 
$\ceff$,  are  the same as in the $b\to s$ analogue, and can be found in \rfs{burev, bm, cella, misiakerr}. 
Using the parameters given in Appendix \ref{app1}, we obtain in leading logarithmic approximation 
\bea\label{wilsonc7c10}
\cseff = -0.315, \quad c_{10} =  -4.642\ ,
\eea
and in next-to-leading approximation
\bea\label{wilsonc9}
\lefteqn{\ceff=c_9 + 0.124\ \omega(\sh) + g(\mch,\sh) \left(3 c_1 + c_ 2 +3c_3 + c_4 + 3c_5 + c_6\right)}\nnu\\ 
& &\mbox{}+ \l_u\left( g(\mch,\sh)- g(\muh,\sh)\right)\left(3 c_1 + c_ 2 \right)  -\frac{1}{2} g(\mdh,\sh) \left(c_3 + 3 c_4\right)
\nnu\\
& & \mbox{}-\frac{1}{2} g(\mbh,\sh)\left(4c_3 + 4c_4 +3c_5+c_6\right)+ 
\frac{2}{9}\left(3c_3 +c_4+3c_5+c_6\right)\ ,
\eea
with
\bea\label{wilsoncoeff}
c_1 &=& -0.249, \quad c_2 = 1.108, \quad c_3 = 1.112\times 10^{-2}, \quad c_4 = -2.569\times 10^{-2}\ ,
\nnu\\
c_5 &=& 7.404\times 10^{-3}, \quad c_6 = -3.144\times 10^{-2}, \quad c_9 = 4.227\ ,
\eea
and the notation $\sh = q^2/m_b^2$, $\mh_q = m_q/m_b$. In the above formula $\omega(\sh)$ 
represents the one-gluon correction to the matrix element of the operator ${\cal O}_9 $ 
(see Appendix \ref{functions}), while the function $g(\mh_q, \sh)$ arises from the one-loop 
contributions of the four-quark operators ${\cal O}_1$--${\cal O}_6$, i.e.
\bea\label{loopfunc}
g(\mh_q,\sh)&=&-\frac{8}{9}\ln(\mh_q)+\frac{8}{27}+\frac{4}{9}y_q
-\frac{2}{9}(2+y_q)\sqrt{|1-y_q|}\nnu\\
&&\times\left\{
\Theta(1-y_q)(\ln\left(\frac{1 + \sqrt{1-y_q}}{1 - \sqrt{1-y_q}}\right)-i\p)
+ \Theta(y_q-1) 2\arctan\frac{1}{\sqrt{y_q-1}}\right\},\nnu\\
\eea
with $y_q \equiv 4 \mh_q^2/\sh$.
\vspace{1.5cm}

\Sec{LONG-DISTANCE EFFECTS: \bm $\r$, \bm $\o$ AND THE \bm $J/\psi$ FAMILY}
A more complete  analysis of the above decay has to take into account long-distance contributions, which 
have their origin in real $u\bar{u}$, $d\bar{d}$, and $c\bar{c}$ intermediate states, i.e.~$\r$, $\o$, and 
$J/\psi$, $\psi'$ etc.,  in addition to the short-distance interaction defined by \eqs{matrixele}{wilsoncoeff}. 
In the case of  the $J/\psi$ family this is usually accomplished  by introducing a 
Breit-Wigner distribution for the resonances through the replacement \cite{longdist}
\be\label{long-dist}
g(\mch,\sh) \longrightarrow g(\mch,\sh)-\frac{3\p}{\a^2}\sum\limits_{V= J/\psi, \psi', \dots}
\frac{\mh_V {\branch}(V\to l^+l^-)\hat{\G}_{\mathrm {total}}^V}
{\sh-\mh^2_V + i \mh_V \hat{\G}_{\mathrm {total}}^V}\ ,
\ee
where the properties of the vector mesons are listed in \rf{pdg}.

We prefer to follow a different procedure, discussed in our previous paper \cite{ks}, which uses the renormalized  photon vacuum polarization 
$\Pi^{\g}_{\mathrm{had}}(\sh)$, related to the  measurable quantity 
$\rhadron(\sh)\equiv \sigma_{\mathrm{tot}}(e^+e^-\to{\mathrm{hadrons}})
/\sigma(e^+e^-\to\m^+\m^-)$. This allows us to implement the long-distance contributions 
 using experimental data. The absorptive part of the vacuum polarization is given by
\bea\label{absorptive}
\Im \Pi^{\g}_{\mathrm{had}}(\sh) = \frac{\a}{3} \rhadron(\sh)\ ,
\eea
whereas the dispersive part may be obtained via a once-subtracted dispersion relation \cite{dispersion}
\be\label{dispersive}
\Re \Pi^{\g}_{\mathrm{had}}(\sh) = \frac{\a \sh}{3\p}\ P \int\limits_{4\hat{m}_{\p}^2}^{\infty}
\frac{R_{\mathrm {had}}(\sh')}{\sh'(\sh'-\sh)}d\sh'\ , 
\ee
where $P$ denotes the principal value. 

To derive an expression that relates $g(\hat{m}_q, \sh)$ and $\rhadron(\sh)$, let us start with the electromagnetic current involving $u$, $d$ and $c$ quarks, which is relevant to the production of $\r$, $\o$ and $J/\psi$ resonances:
\bea\label{emcurrent}
j_{\m}^{\mathrm{em}} = \frac{2}{3} \bar{u} \g_{\m} u - \frac{1}{3} \bar{d} \g_{\m} d
+ \frac{2}{3} \bar{c} \g_{\m} c + \cdots\ .
\eea
Using \eq{emcurrent}, the vacuum polarization may then be written as
\bea\label{vacpola}
\Pi^{\g}_{\mathrm{had}} = \frac{4}{9}\Pi^{c\bar{c}} + \frac{4}{9} \Pi^{u\bar{u}} 
+ \frac{1}{9} \Pi^{d\bar{d}} + \cdots  \ .
\eea
 The vacuum polarization $\Pi^{q\bar{q}}$ associated with a $q\bar{q}$ loop is related to 
$g(\hat{m}_q, \sh)$ via
\bea
\Pi^{q\bar{q}} = \frac{9}{4} \frac{\a}{\p} \left (g(\hat{m}_q, \sh) + \frac{4}{9} 
+ \frac{8}{9} \ln \hat{m}_q\right) .
\eea
Next we define currents corresponding to the quantum numbers of $\r$, $\o$, and $J/\psi$
\bea\label{mesonstates}
j_{\m}^{\r} = \frac{1}{2}( j_{\m}^u - j_{\m}^d), \quad 
j_{\m}^{\o} = \frac{1}{6}( j_{\m}^u + j_{\m}^d), \quad 
j_{\m}^{J/\psi} = \frac{2}{3} j_{\m}^c\ ,
\eea
with $j_{\m}^q = \bar{q}\g_{\m}q$, in terms of which the vacuum polarization, \eq{vacpola},  can be 
rewritten as
\bea
\Pi^{\g}_{\mathrm{had}} = \Pi^{J/\psi} + \Pi^{\o} + \Pi^{\r} + \cdots \ .
\eea
With the assumption $m_u = m_d$ it follows immediately that $\Pi^{u \bar{u}} = \Pi^{d\bar{d}}$, and 
we arrive at
\bea
\Pi^{c\bar{c}} = \frac{9}{4} \Pi^{J/\psi}\ ,
\eea
\bea
\Pi^{u\bar{u}} = \Pi^{d\bar{d}} = \frac{9}{5} (\Pi^{\o} + \Pi^{\r})\ ,
\eea
so that
\bea\label{relaccim}
\Im g(\mch, \sh) = \frac{\p}{3}\rhadronpsi\ ,
\eea 
\bea\label{relauuim}
\Im g(\muh, \sh) = \Im g(\mdh, \sh) = \frac{4\p}{15} (\rhadronrho + \rhadronomega)\ .
\eea
For the real part of the one-loop function $g(\hat{m}_q, \sh)$ one finds
\bea\label{relaccrel}
\Re  g(\mch,\sh) = -\frac{8}{9} \ln \mch -\frac{4}{9} + \frac{\sh}{3}\ P \int\limits_{4\hat{m}_{D}^2}^{\infty}
\frac{R_{\mathrm {had}}^{J/\psi}(\sh')}{\sh'(\sh'-\sh)}d\sh'\ ,
\eea
and 
\bea\label{relauurel}
\Re  g(\hat{m}_q,\sh) = -\frac{8}{9} \ln \hat{m}_q -\frac{4}{9} + \frac{4 \sh}{15}\ 
P \int\limits_{4\hat{m}_{\p}^2}^{\infty}
\frac{R_{\mathrm {had}}^{\r}(\sh')+ R_{\mathrm {had}}^{\o}(\sh')}{\sh'(\sh'-\sh)}d\sh',
\ q = u, d\ .
\eea 
Note that in many cases the evaluation of the dispersion integral may be carried out 
analytically (see e.g.~\rf{fred}).
The cross-section ratios appearing in \eqs{relaccim}{relauurel} may be written as 
\bea
\rhadronpsi = \rcontcc + \rrespsi\ ,
\eea
\bea
\rhadronrho + \rhadronomega = \rcontuu + \rresrho+\rresomega\ ,
\eea
where the subscripts ``cont'' and ``res'' refer to the contributions from the continuum and the resonances
respectively. The $J/\psi$ resonances and  $\o$ are well described through a relativistic Breit-Wigner 
form, i.e. 
\be\label{BWpsi}
\rrespsi  =\sum\limits_{V= J/\psi, \psi', \dots}
 \frac{9\sh}{\a^2}\ \frac{{\branch}(V\to l^+l^-)\hat{\G}_{\mathrm {total}}^{V}
\hat{\G}_{\mathrm {had}}^{V}}
{(\sh-\mh^2_V)^2 +  \mh_V^2 \hat{\G}_{\mathrm {total}}^{V^2}}\ , 
\ee
and
\be\label{BWomega}
\rresomega  =
\frac{9\sh}{\a^2}\ \frac{{\branch}({\o}\to l^+l^-)\hat{\G}_{\mathrm {total}}^{{\o}^2}}
{(\sh-\mh^2_{\o})^2 +  \mh_{\o}^2 \hat{\G}_{\mathrm {total}}^{{\o}^2}}\ , 
\ee 
with a $\sh$-independent total width, which is quite adequate for our purposes. 
The $\r$ resonance may be introduced through 
\bea
\rresrho = \frac{1}{4} \left(1-\frac{4\hat{m}_{\p}^2}{\sh}\right)^{3/2} |F_{\p}(\sh)|^2\ ,
\eea
$F_{\p}(\sh)$ being the pion form factor, which is represented by a modified Gounaris-Sakurai formula
\cite{gounsak}. The continuum contributions can be parametrized using the experimental data from
\rf{bp}, and are given in Appendix \ref{app1}. 
\vspace{1.5cm}

\Sec{BRANCHING RATIO AND\cpbf-VIOLATING ASYMMETRY}
The differential branching ratio for $b\to d \,l^+ l^-$ in the variable $\sqrt{\sh}$ including next-to-leading order QCD corrections is given by
\be\label{rate}
\frac{d\,{\branch}}{d \sqrt{\sh}} = \frac{\a^2}{2\p^2}\,\frac{|V_{tb}^{}V_{td}^*|^2}{|V_{cb}|^2}\,
\frac{{\branch} (B\to X_c e \bar{\n}_e)}{f(\mch) \k (\mch)}\,
\l^{1/2}(1,\sh,\mdh^2) \,\sqrt{\sh-4 \mlh^2}\,\Sigma\ ,
\ee
where we have neglected nonperturbative corrections of $O(1/m_b^2)$ \cite{fls}.
The various factors appearing in \eq{rate} are  defined by
\bea
\l (a,b,c) &=& a^2 + b^2 + c^2 - 2 (a b + b c + a c)\ ,\\
\Sigma &=& \Bigg\{\left(12\, \Re(\cseff \ceff)F_1(\sh,\mdh^2) + \frac{4}{\sh}|\cseff|^2 F_2(\sh,\mdh^2)\right)
\left(1+\frac{2\mlh^2}{\sh}\right)\nnu \\
&&+\left(|\ceff|^2 + |c_{10}|^2\right) F_3(\sh,\mdh^2,\mlh^2)+ 6 \mlh^2\left(|\ceff|^2 
-|c_{10}|^2\right) F_4(\sh,\mdh^2)\Bigg\}\ ,\nnu\\ \label{Delta}
\eea
with

\parbox{12cm}{
\bea\label{kinfuncs}
F_1(\sh,\mdh^2)&=&(1-\mdh^2)^2 - \sh(1+\mdh^2)\ ,\nnu\\
F_2(\sh,\mdh^2)&=& 2 (1+\mdh^2) (1-\mdh^2)^2 -\sh (1 + 14 \mdh^2 +\mdh^4)-\sh^2 (1+\mdh^2)\ ,\nnu\\ 
F_3(\sh,\mdh^2,\mlh^2) &=& (1-\mdh^2)^2 +\sh (1+\mdh^2) -2 \sh^2 
+ \l(1,\sh,\mdh^2) \frac{2\mlh^2}{\sh}\ ,\nnu\\
F_4(\sh,\mdh^2)&=&1- \sh +\mdh^2\ ,\nnu
\eea}\hfill\parbox{0.8cm}{\bea\eea}
while the ratio of CKM matrix elements  in terms of the Wolfenstein parameters $\r$ and $ \eta$ has 
already been given in \eq{ckmwolf1}. In order to remove the uncertainties in \eq{rate} due to an 
overall factor of $m_b^5$, we have introduced the inclusive semileptonic branching ratio via the relation
\bea\label{semilept}
\G (B\to X_c e \bar{\n}_e) = \frac{G_F^2 m_b^5}{192 \p^3} |V_{cb}|^2 f(\mch) \k (\mch)\ ,
\eea  
where $f(\mch)$ and $\k (\mch)$ represent the phase space and the one-loop QCD corrections \cite{maiani} 
to the semileptonic decay respectively, and are given in Appendix \ref{functions}. 
Integrating the distribution in \eq{rate} for $l = e$, $\m$, and $\t$ over $\sqrt{\sh}$, we obtain the 
branching ratio ${\mathrm{Br}}\,(B\to X_d\, l^+ l^-)$, depending on the specific choice of $\r$ and $\eta$. 
The results are shown in Table \ref{table1}, for typical  values of $(\r, \eta)$ in the experimentally allowed 
domain \cite{alirev}.\bem{The  branching ratio for different regions of $\sqrt{\sh}$ will be discussed below.} 
Note that the branching ratio is quite sensitive to the Wolfenstein parameter $\r$. For instance, the 
branching ratio for $B\to X_d\, e^+ e^-$ varies from  $2.7$ to $7.9 \times 10^{-7}$, when $\r$ is varied from 
$+0.3$ to $-0.3$.
%
%
\begin{table}
\begin{center}
\caption{Branching ratio $\protect {\mathrm{Br}}\,(B\to X_d\, l^+ l^-)$, where $l = e$, $\m$ or $\t$, 
for different values of $(\r, \eta)$ excluding the region $(\pm 20\ \MeV)$ around  the $J/\psi$  and $\psi'$ 
resonances.}\label{table1}
\vspace{0.3cm}
\begin{tabular}{cccc}
\hline\hline
  \multicolumn{1}{c}{$(\rho, \eta)$}  &  \multicolumn{1}{c}{${\mathrm{Br}}\,
(B\to X_d\, e^+ e^-)$} & \multicolumn{1}{c}{${\mathrm{Br}}\,
(B\to X_d\, \m^+ \m^-)$} &\multicolumn{1}{c}{${\mathrm{Br}}\,
(B\to X_d\, \t^+ \t^-)$}  \\ \hline
$(0.3, 0.34)$ & $2.7\times 10^{-7}$& $1.8\times 10^{-7}$&$0.7\times 10^{-8}$\\
$(-0.07, 0.34)$ &$5.5\times 10^{-7}$&$3.8\times 10^{-7}$&$1.6\times 10^{-8}$ \\
$(-0.3, 0.34)$ &$7.9\times 10^{-7}$&$5.4\times 10^{-7}$ & $2.3\times 10^{-8}$\\
\hline\hline
\end{tabular}
\end{center}
\end{table}
%
%

Let us now turn to the \cp-violating rate asymmetry, which is defined as follows: 
\bea\label{asym}
\asym(\sqrt{\sh}) = \frac{d \G/d\sqrt{\sh} - d \ol{\G}/d\sqrt{\sh}}{d \G/d\sqrt{\sh} 
+ d \ol{\G}/d\sqrt{\sh}}\ ,
\eea
where
\bea
\frac{d \G}{d \sqrt{\sh}} \equiv  \frac{d \G (b\to d\,l^+l^-)}{d \sqrt{\sh}},
\quad \quad \frac{d\ol{\G}}{d \sqrt{\sh}} \equiv  \frac{d\G (\bar{b} \to \bar{d}\, l^+l^-)}{d \sqrt{\sh}}\ .
\eea
The physical origin of a \cp-violating asymmetry in the reaction can be understood by considering the 
term proportional to $\ceff$ in the matrix element, which can be written symbolically as
\bea\label{particle}
{\cal{M}} \sim  A +  \l_u B\ .
\eea
The corresponding matrix element for $\bar{b}\to\bar{d}\,l^+l^-$ is
\bea\label{antiparticle}
\ol{{\cal{M}}} \sim A + \l_u^* B\ ,
\eea
giving an asymmetry 
\bea\label{asymAB}
\asym = \frac{-2\,\Im \l_u \Im (A^* B)}{|A|^2 + |\l_u B|^2 + 2 \Re \l_u\Re (A^* B)}\ , 
\eea
which provides a measure for \cp\ violation. The asymmetry results from the presence of \cp\ violation in the 
CKM matrix ($\Im \l_u \neq 0$) and unequal unitarity phases in the amplitudes $A$ and 
$B$ ($\Im (A^* B) \neq 0$).  

The complete result contains an additional term due to the  interference of $\cseff$ with  $\ceff$,
and the asymmetry takes the final form
\bea\label{asymcomp}
\asym(\sqrt{\sh}) = \frac{ -2 \Im \l_u \Delta}{\Sigma + 2 \Im \l_u \Delta}\approx -2 \Im \l_u \frac{\Delta}{\Sigma}
= \left(\frac{2\eta}{(1-\r)^2 + \eta^2}\right) \frac{\Delta}{\Sigma}\ ,
\eea
with $\Sigma$ defined in \eq{Delta}, and 
\bea
\ceff &\equiv& \xi_1 + \l_u \xi_2\ ,\nnu\\
\Delta &=& \Im (\xi_1^* \xi_2^{})f_+(\sh) + \Im (\cseff \xi_2^{}) f_1(\sh)\ ,\nnu\\
f_1(\sh)&=& 6 F_1(\sh, \mdh^2) \left(1+\frac{2\mlh^2}{\sh}\right)\ ,\nnu\\
f_+(\sh)&=&  F_3(\sh, \mdh^2, \mlh^2) + 6 \mlh^2 F_4(\sh, \mdh^2)\ ,
\eea
where the phase-space functions $F_1$ and $F_{3, 4}$ are given in \eq{kinfuncs}. 
Notice that $\asym$  vanishes  as $m_u\to m_c$, since in that limit $\xi_2 \to 0$ (see \eq{wilsonc9}).

\begin{figure}[h]
\vskip -3cm
\centerline{\epsfysize=14cm\epsffile{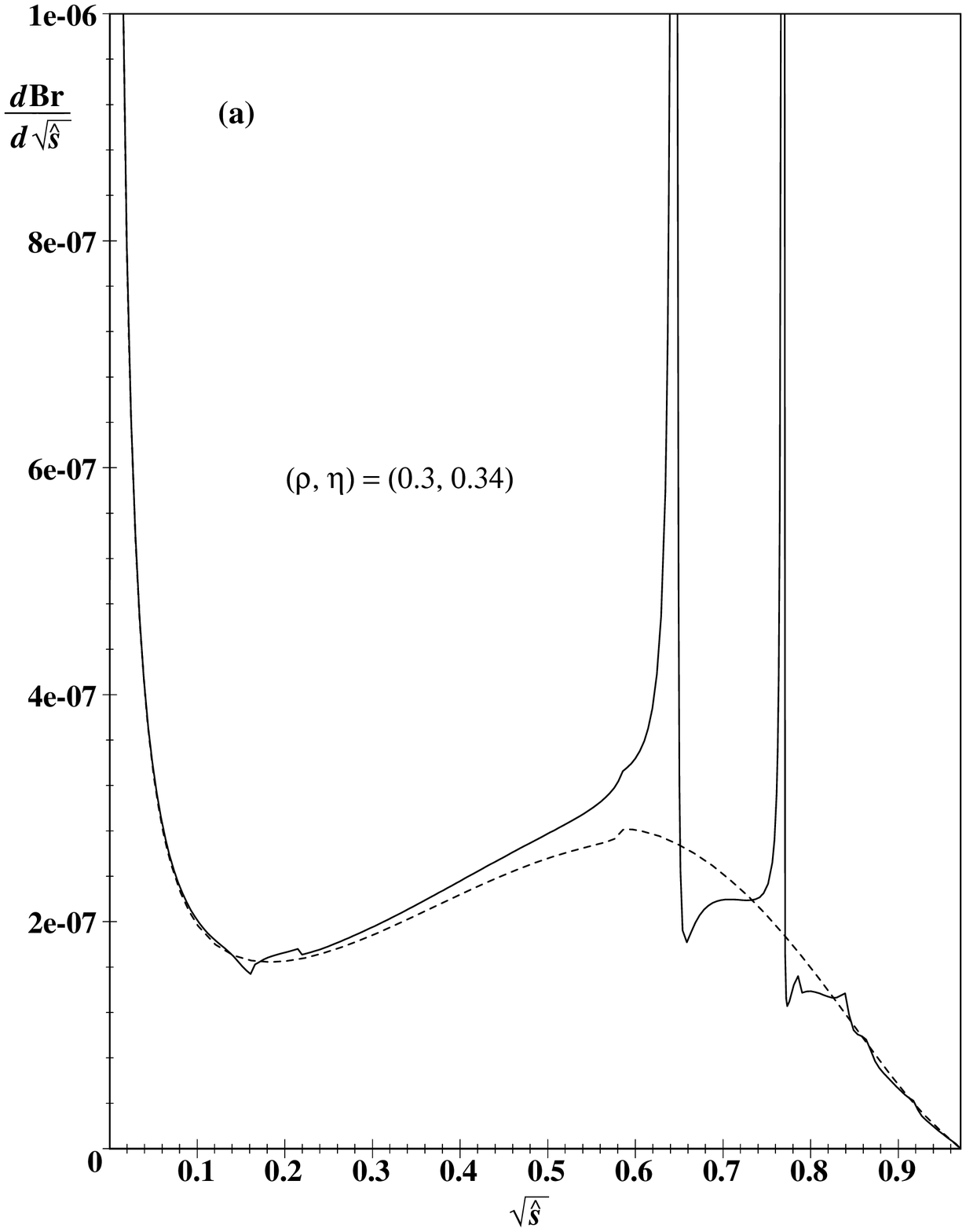}}
\vskip -2cm
\centerline{\epsfysize=14cm\epsffile{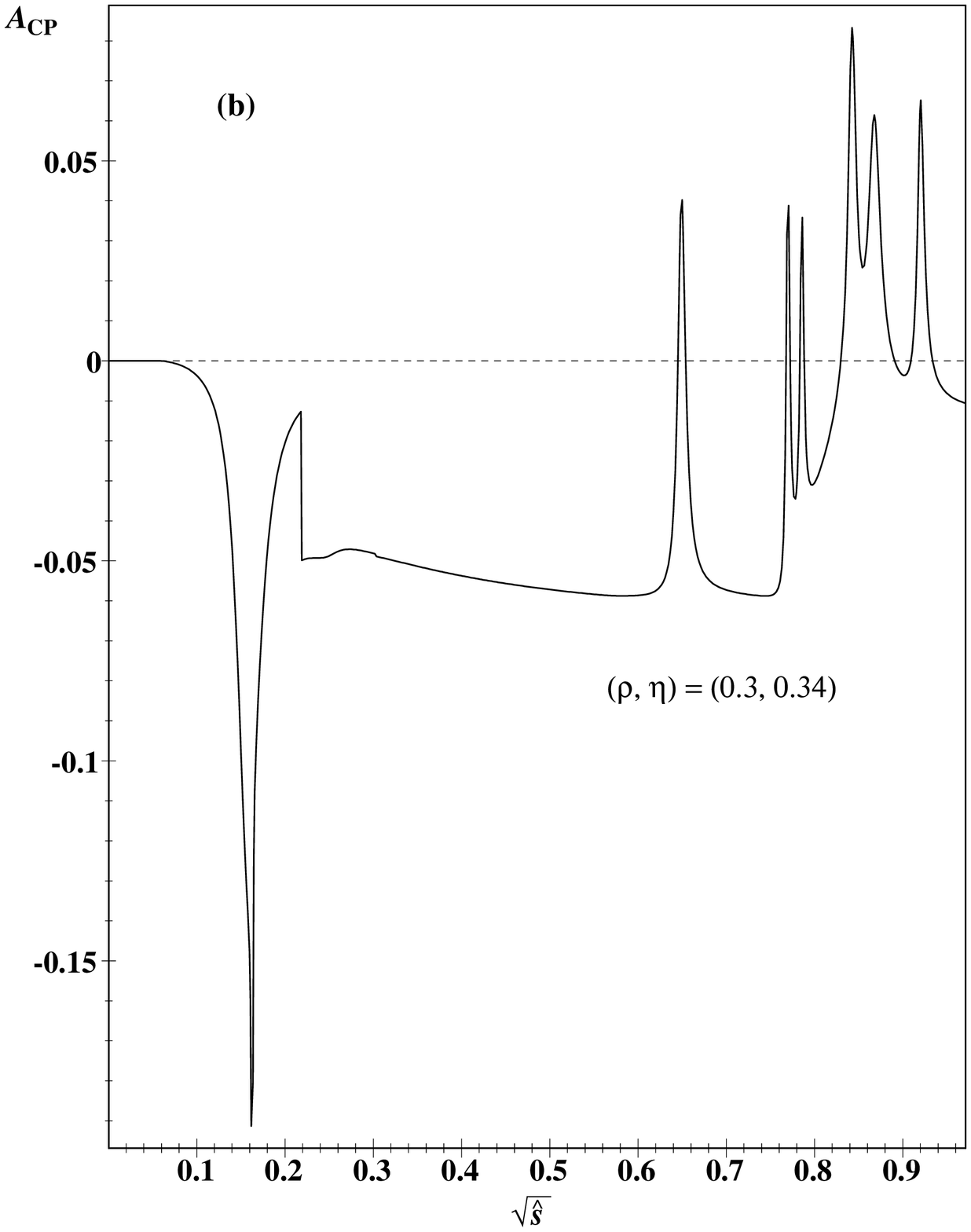}}
\caption{}\label{fig1}
\end{figure}

\begin{figure}[h]
\vskip -3cm
\centerline{\epsfysize=14cm\epsffile{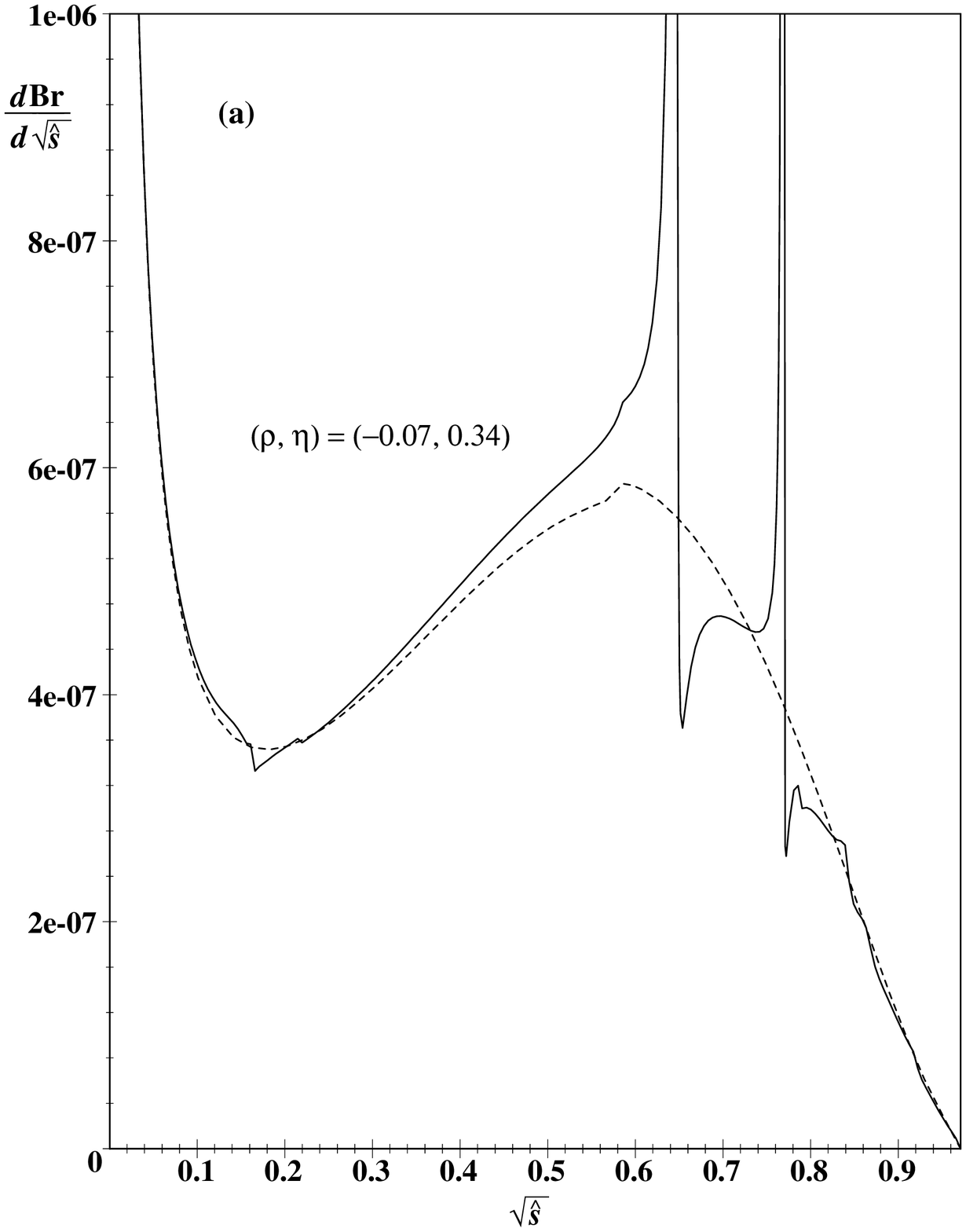}}
\vskip -2cm
\centerline{\epsfysize=14cm\epsffile{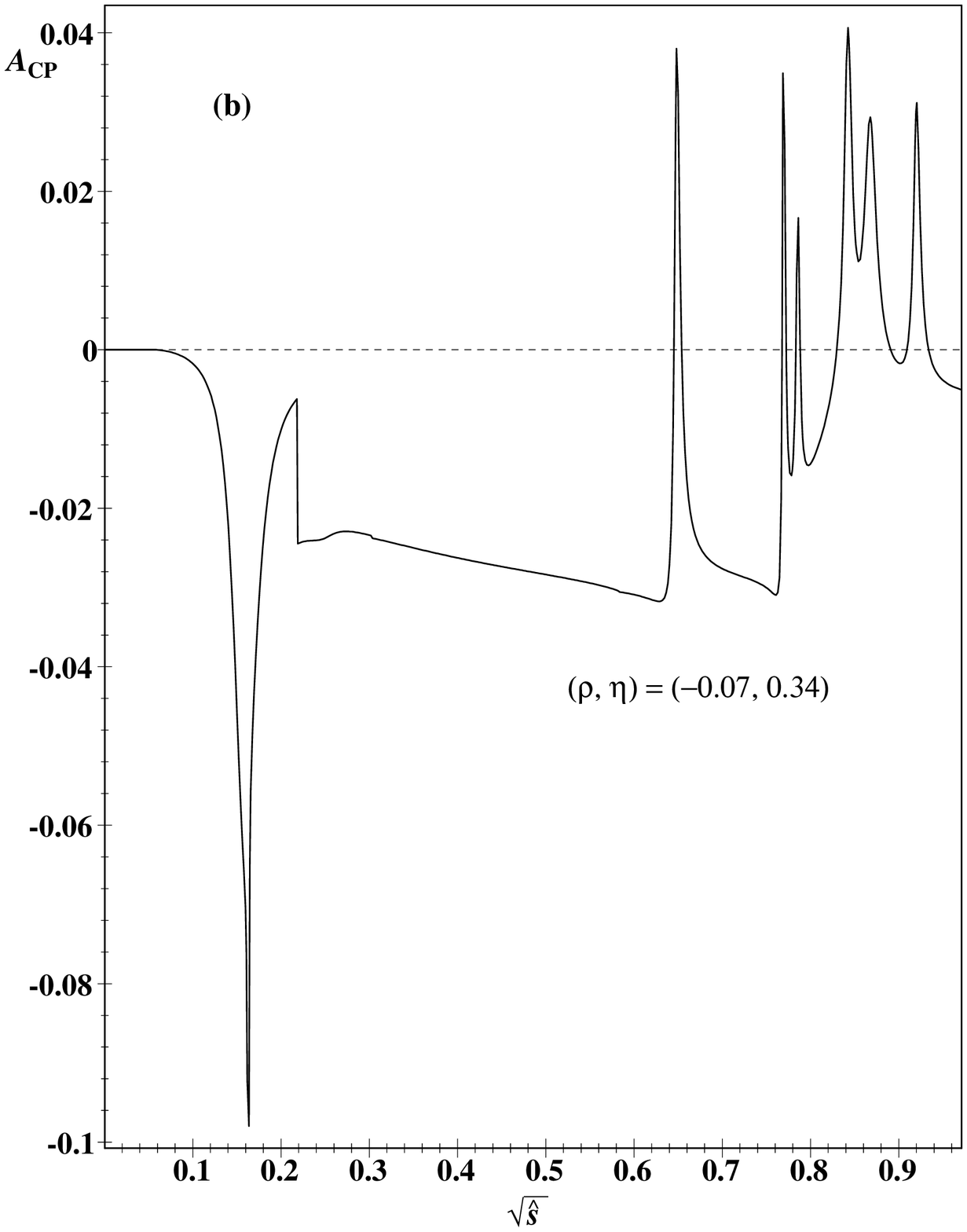}}
\caption{}\label{fig2}
\end{figure}

\begin{figure}[h]
\vskip -3cm
\centerline{\epsfysize=14cm\epsffile{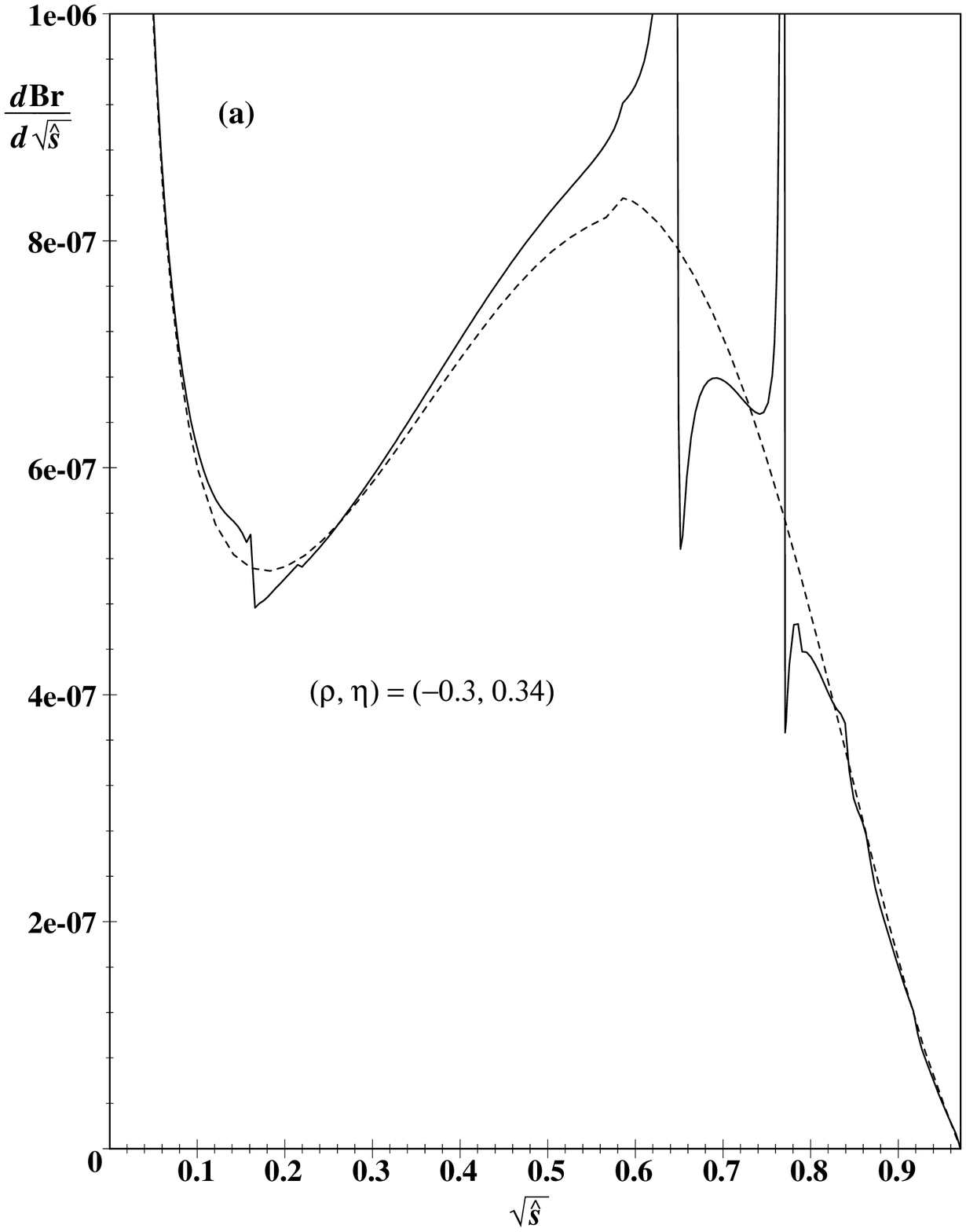}}
\vskip -2cm
\centerline{\epsfysize=14cm\epsffile{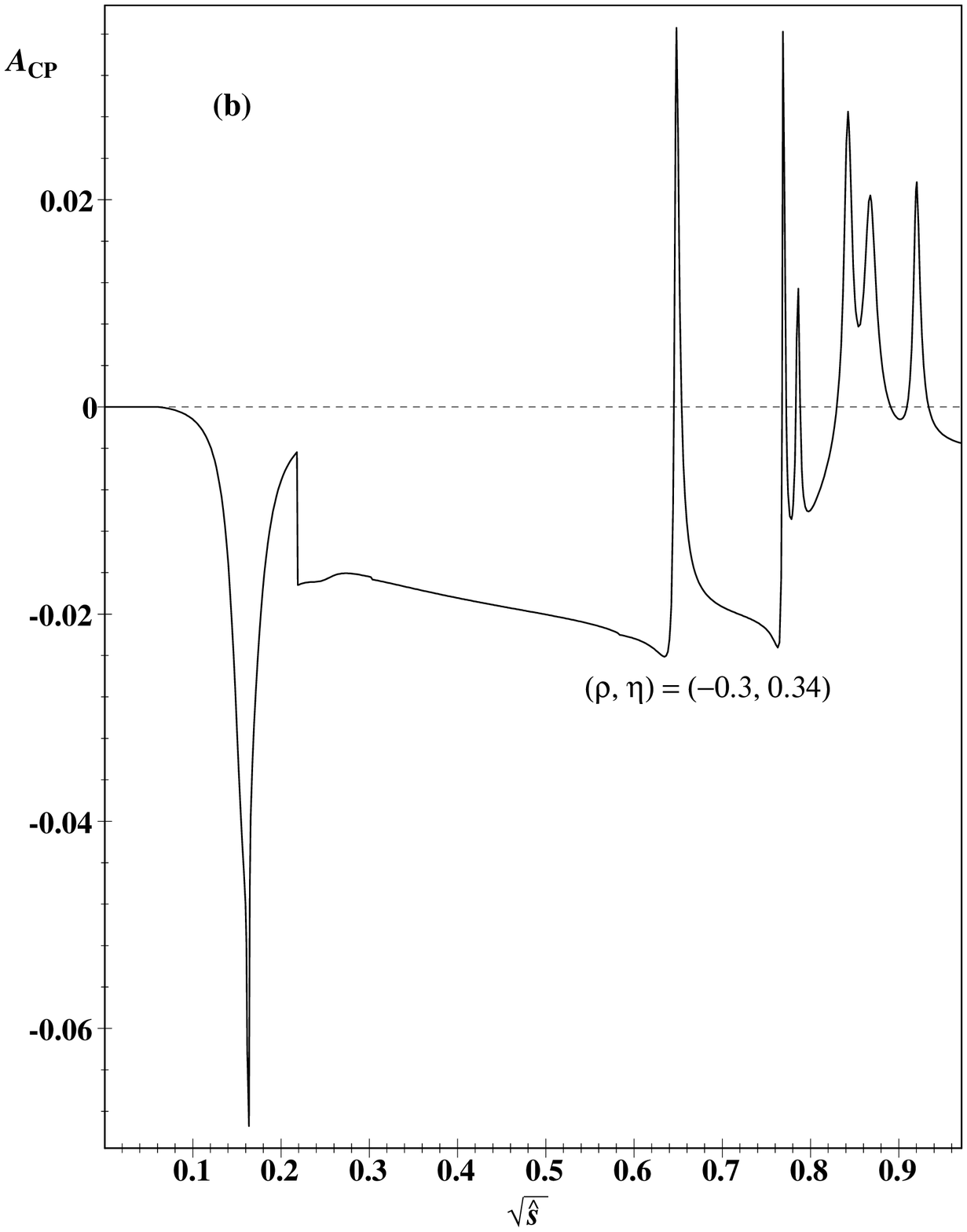}}
\caption{}\label{fig3}
\end{figure}

Our numerical  results for the asymmetry together with the differential branching ratio, \eq{rate}, 
are shown  in Figs.~\ref{fig1}--\ref{fig3} for different values of $\r$ and $\eta$.\bem{We have also calculated the asymmetry in the $b \to s$ transition, which is roughly one order of magnitude smaller than in $b\to d$. 
Our results for the asymmetry differ somewhat from those given in \rf{duyang}, which uses an incorrect 
sign for the absorptive part of  the one-loop function $g(\hat{m}_q, \sh)$.
The correct sign is given in Refs.~\cite{bm} and \cite{misiakerr}.} 
It is interesting to note that the $\r$ resonance is barely visible in the invariant mass spectrum, but has a 
strong influence on the  asymmetry in the region up to 1 GeV. We have
evaluated the branching ratio and average asymmetry $\left\langle \asym \right\rangle$ for different
regions of $\sqrt{s}$ using \eq{asym}, and our results are displayed in 
Tables \ref{table2}--\ref{table4}.\bem{A variation of $m_t$ in the interval $176\pm 10\ \GeV$ changes 
these numbers by $\lesssim 10 \%$.}  
%
%
\begin{table}
\begin{center}
\caption{Branching ratio $\protect {\mathrm{Br}}\,(B\to X_d\, e^+ e^-)$ and average asymmetry  
$\protect\left\langle\asym\right\rangle$ 
for different regions of $\protect\sqrt{s}$, below the $J/\psi$ 
resonance $(\varepsilon =  20\,\MeV)$.}\label{table2}
\vspace{0.3cm}
\begin{tabular}{cccc}
\hline\hline
 &  \multicolumn{1}{c}{$(\rho, \eta)$}  &  \multicolumn{1}{c}{$2m_e < \sqrt{s} < 1\,\GeV $}  &
\multicolumn{1}{c}{$1\,\GeV < \sqrt{s} < (m_{J/\psi} - \varepsilon)$} \\ \hline
${\mathrm{Br}}$&$(0.3, 0.34)$ &$1.1\times 10^{-7}$ & $1.2\times 10^{-7}$ \\ 
&$(-0.07, 0.34)$  &$2.4\times 10^{-7}$ &$2.3\times 10^{-7}$  \\ 
&$(-0.3, 0.34)$ &$3.4\times 10^{-7}$ & $3.3\times10^{-7}$\\ 
\hline
$\left\langle\asym\right\rangle$ & $(0.3, 0.34)$&$-8.4\times 10^{-3}$ &$-5.3\times 10^{-2}$  \\ 
&$(-0.07, 0.34)$  &$- 4.0\times 10^{-3}$ &$-2.7\times 10^{-2}$   \\ 
 & $(-0.3, 0.34)$&$ -2.9\times 10^{-3}$& $-1.9\times 10^{-2}$  \\ 
\hline\hline
\end{tabular}
\end{center}
\end{table}
%
%
\begin{table}
\begin{center}
\caption{Branching ratio $\protect {\mathrm{Br}}\,(B\to X_d\, e^+ e^-)$ and average asymmetry  
$\protect\left\langle\asym\right\rangle$ for the large $\protect \sqrt{s}$ region, excluding the 
$\protect J/\psi$ and $\protect \psi'$ resonances  $\protect (\varepsilon= 20\,\MeV)$.}\label{table3}
\vspace{0.3cm}
\begin{tabular}{cccc}
\hline\hline
 &  \multicolumn{1}{c}{$(\rho, \eta)$}  &  \multicolumn{1}{c}{$(m_{J/\psi} + \varepsilon)
< \sqrt{s} < (m_{\psi'} - \varepsilon) $}  &
\multicolumn{1}{c}{$(m_{\psi'} + \varepsilon)< \sqrt{s} < {\sqrt{s}}_{\mathrm{max}}$} \\ \hline
${\mathrm{Br}}$&$(0.3, 0.34)$ &$0.3\times 10^{-7}$ & $1.6\times 10^{-8}$ \\ 
&$(-0.07, 0.34)$  &$0.5\times 10^{-7}$ &$3.4\times 10^{-8}$  \\ 
&$(-0.3, 0.34)$ &$0.8\times10^{-7}$ &$4.9\times 10^{-8}$ \\ 
\hline
$\left\langle\asym\right\rangle$ & $(0.3, 0.34)$&$-5.1\times 10^{-2}$ &$5.2\times 10^{-3}$  \\ 
&$(-0.07, 0.34)$  & $-2.5\times 10^{-2}$&$2.1\times 10^{-3}$   \\ 
 & $(-0.3, 0.34)$& $-1.8\times 10^{-2}$& $1.5\times 10^{-3}$ \\ 
\hline\hline
\end{tabular}
\end{center}
\end{table}
%
%
\begin{table}
\begin{center}
\caption{Branching ratio $\protect {\mathrm{Br}}\,(B\to X_d\, e^+ e^-)$ and average asymmetry  
$\protect\left\langle\asym\right\rangle$ near the $J/\psi$  and $\psi'$ resonances 
($ \varepsilon = 20\,\MeV$).}\label{table4}
\vspace{0.3cm}
\begin{tabular}{ccc}
\hline\hline
  &  \multicolumn{1}{c}{$(m_{J/\psi} - \varepsilon)
< \sqrt{s} < (m_{J/\psi} + \varepsilon) $}  &
\multicolumn{1}{c}{$(m_{\psi'} -\varepsilon)< \sqrt{s} < (m_{\psi'} +\varepsilon)$} \\ \hline
${\mathrm{Br}}$&$3.7\times 10^{-6}$ & $1.8\times 10^{-7}$ \\ 
$\left\langle\asym\right\rangle $ &$0.6\times 10^{-3}$ & $4.4\times 10^{-3}$ \\ 
$\left\langle\asym\right\rangle^{\,\mathrm a}$ &$2.9\times 10^{-3}$ & $6.7\times 10^{-3}$ \\ 
\hline\hline
\multicolumn{3}{l}{$^{\mathrm a}$ Including OZI correction, induced by one-photon exchange as specified
in \eq{ozi}.}
\end{tabular}
\end{center}
\end{table}
%
%
\newpage
\Sec{CONCLUSIONS}\label{comments}
The principal results of our analysis are as follows:
\ben
\renewcommand{\labelenumii}{({\roman{enumii}})}
\item In the region excluding the $J/\psi$ resonances, we find a sizeable \cp-violating asymmetry between the decays $b\to d\, e^+e^-$ and $\bar{b}\to \bar{d}\, e^+e^-$. This asymmetry amounts to $-5.3\%$ ($-1.9\%$)
for the invariant mass region $1\ \GeV < \sqrt{s}_{e^+ e^-} < m_{J/\psi} -20\ \MeV$, assuming $\eta = 0.34$ and $\r = 0.3$ ($-0.3$). The corresponding branching ratio is $1.2\times 10^{-7}$ ($3.3\times 10^{-7}$).
The asymmetry scales approximately as $\eta \left( (1-\r)^2 + \eta^2\right)^{-1}$, while the branching ratio scales as $(1-\r)^2 + \eta^2$. For a nominal asymmetry of $5\%$ and a branching ratio of $10^{-7}$, a measurement at $3\s$ level requires $4\times 10^{10}$ $B$ mesons. In view of the clear dilepton signal, such a measurement might be feasible at future hadron colliders. It should be noted, however, that identification of the reaction $b\to d\, e^+e^-$ in the presence of the much stronger reaction $b\to s\, e^+e^-$ would require a study of the decay vertex, in order to select final states such as $\pi^+$, $\pi^+\pi^-\pi^+$ etc.~(accompanied by any numbers of neutrals). In the inclusive analysis of $e^+e^-$ pairs, only those with invariant mass in the range 
$(M_B - M_K) < \sqrt{s} < (M_B - M_{\pi})$ can be unambiguously ascribed to $b\to d\, e^+e^-$.
\item In the neighbourhood of the $J/\psi$ resonance $(m_{J/\psi} -20\ \MeV <  \sqrt{s} < m_{J/\psi} +20\ \MeV)$,
the branching ratio is substantial $(\branch = 3.7\times 10^{-6})$, but the asymmetry is very small
$(\big\langle \asym^{J/\psi}\big\rangle = 0.6\times 10^{-3})$. This smallness in asymmetry is the inevitable result of a very large 
$c\bar{c}$ amplitude near the $J/\psi$, interfering with a small nonresonant background.
\item It is pertinent to ask if some refinement of the effective Hamiltonian underlying our calculation might lead to a higher asymmetry in the $J/\psi$ region. In this connection, the following comments are in order.
\ben
\item  Our prescription for incorporating resonances into the effective Hamiltonian via the vacuum polarization
function $\Pi^{\g}_{\mathrm{had}}(\sh)$ implicitly assumes that the transition $b\to dJ/\psi$ is adequately described by the leading term $(3 c_1 + c_2+ 3c_3+c_4+3c_5+c_6)$ appearing in the Wilson coefficient $\ceff$, 
\eq{wilsonc9}. With the values
of $c_i$ $(i = 1, \dots, 6)$ given in \eq{wilsoncoeff}, the theoretical branching ratio for the related reaction
 $b\to sJ/\psi$ is known to be $\sim 5$ times smaller than measured \cite{alirev, ks, lw}. 
It could be argued that for the purposes of calculating the $b\to d J/\psi$ amplitude the coefficient 
$(3 c_1 + c_2+\cdots)$ should accordingly be corrected to 
$\k_V (3 c_1 + c_2+\cdots)$, with $\k_V \sim \sqrt{5}$. 
While such a procedure {\it enhances\/} the branching 
ratio of $b\to dJ/\psi$ by a factor $\k_V^2$, it {\it reduces\/} the asymmetry by a factor ${\k_V}$. 
Outside the $J/\psi$ and $\psi'$ regions, the branching ratio is essentially independent of $\k_V$. The asymmetry for $\sqrt{s} < m_{J/\psi}$ is likewise unaffected, while that between $J/\psi$ and $\psi'$ is reduced by 
$\sim \k_V$. In the region $\sqrt{s} > m_{\psi'}$ the asymmetry is quite sensitive to $\k_V$ and can even be enhanced by an order of magnitude. This corner of phase space accounts, however, for only about $6 \%$ of the decay rate.
\item
The asymmetry may be slightly enhanced if one takes into account mixing of the $c\bar{c}$ current with the 
$u\bar{u}$ and $d\bar{d}$ currents. Such a mixing can give rise to an OZI-rule violating transition
$u\bar{u} \to J/\psi$, mediated by a one-photon (or 3-gluon) intermediate state \cite{soadun, soares}. The QED effect
can be incorporated into our calculation of the asymmetry near the $J/\psi$ resonance  by the replacement
\bea\label{ozi}
 \l_u g(\mch, \sh) \longrightarrow \l_u( 1 +i \frac{4}{9} \a) g(\mch, \sh)
\eea
in the coefficient $\ceff$. The resulting asymmetry increases from $ 0.6\times 10^{-3}$ to 
$2.9\times 10^{-3}$ (see Table \ref{table4}).
\item Finally, it is possible to contemplate gluonic corrections to the effective Hamiltonian, that allow the 
transition $b\to dJ/\psi$ to take place not only through a colour-singlet $(c\bar{c})_1$ intermediate state
(i.e.~$b\to d(c\bar{c})_1 \to d J/\psi$) but also through a colour-octet intermediate configuration 
($b\to d (c\bar{c})_8 \to d J/\psi$). An illustrative calculation by Soares \cite{soares} yields an asymmetry
of about $1\%$ from such a mechanism. 
\een
\een
Our general conclusion is that a measurement of the branching ratio and partial width asymmetry in the 
channel $b\to d\,e^+e^-$ in the nonresonant continuum, would provide a theoretically clean and fundamental test of the idea that \cp\ violation originates in the CKM matrix. 
The predicted asymmetry in the region $1\ \GeV < m_{e^+e^-} < m_{J/\psi}$ is approximately
\bea
-5.3\%\left(\frac{\eta}{0.34}\right) \left(\frac{1.2\times 10^{-7}}{\branch}\right),
\eea
where $\mathrm{Br}$ denotes the branching ratio in the above interval.
Measurements near the $J/\psi$ resonance are predicted to show a  very small asymmetry 
$(\sim 3\times 10^{-3})$ that depends somewhat on the manner in which QCD modulates the effective 
interaction for $b\to d J/\psi$. 
\vspace{1.5cm}

\centerline{\bf ACKNOWLEDGMENTS}
One of us (F.\,K.) gratefully acknowledges financial support from the Deutsche Forschungsgemeinschaft
(DFG) through Grant No.~Se 502/4-1.
\newpage
\app{INPUT PARAMETERS}\label{app1}
\vspace{-1.2cm}
\bea
&&m_b=4.8\ \GeV,\ m_c= 1.4\ \GeV,\ m_u=m_d= m_{\pi} = 0.139\ \GeV\ ,\nnu\\
& & m_t=176\ \GeV\ ,\ m_e=0.511 \ \MeV,\ m_{\m} = 0.106\ \GeV,\ m_{\t} = 1.777\ \GeV\ , \nnu\\
&&\m=m_b,\ {\branch} (B\to X_c e \bar{\n}_e)= 10.4\%,\ \l = 0.2205,\ 
\Lambda_{\mathrm{QCD}} = 225\ \MeV\ , \nnu \\
&&\a=1/129,\ {\sin}^2{\theta_{\mathrm{W}}} = 0.23,\ M_W= 80.2\ \GeV\ .
\eea

\bea
\rcontuu = 
\left\{\begin{array}{l} 0 \quad{\mathrm{for}}\quad 0\leq \sh \leq 4.8\times10^{-2}\ ,\\ 
1.67 \quad{\mathrm{for}}\quad 4.8\times10^{-2}\leq \sh \leq 1\ .
\end{array}\right.
\eea

\bea
\rcontcc = 
\left\{\begin{array}{l} 0 \quad{\mathrm{for}}\quad 0\leq \sh \leq 0.60\ ,\\ 
-6.80  + 11.33 \sh\quad{\mathrm{for}}\quad 0.60\leq \sh \leq 0.69\ ,\\
1.02\quad{\mathrm{for}}\quad 0.69 \leq \sh \leq 1\ .
\end{array}\right.
\eea
\vspace{1cm}

\app{USEFUL FUNCTIONS}
\label{functions}
As noted by Misiak \cite{misiakerr}, the function $\omega(\sh)$ can be inferred from \cite{jk} and 
is defined by  
\bea
\omega(\sh) &=&- \frac{2}{9} \p^2 - \frac{4}{3}{\mathrm{Li}}_2(\sh) - \frac{2}{3} \ln \sh \ln (1-\sh)
 - \frac{5 + 4\sh}{3(1+2\sh)}\ln (1-\sh)\nnu\\
& &-\frac{2\sh(1+\sh)(1-2\sh)}{3(1-\sh)^2(1+2\sh)} \ln \sh + \frac{5+9\sh - 6 \sh^2}{6(1-\sh)(1+2\sh)}\ .
\eea

\bea
f(\mch) = 1- 8\mch^2 +8\mch^6 - \mch^8 - 24 \mch^4 \ln \mch\ .  
\eea
\bea
\k (\mch) =  1- \frac{2\a_s(m_b)}{3\p}\left( (\p^2-\frac{31}{4})(1-\mch)^2 +\frac{3}{2}\right)\ .
\eea
%
%

%
%
\newpage
\centerline{\bf FIGURE CAPTIONS}
\begin{enumerate}
\item[\bf Figure 1] Branching ratio ${\mathrm{Br}}\,(B\to X_d\, e^+ e^-)$ (a) and \cp-violating asymmetry
$\asym$ (b) including next-to-leading order QCD corrections as well as long-distance contributions
(solid line),  i.e.~$\r$, $\o$, and the $J/\psi$ family, as a function of $\sqrt{\sh}$, $\sh\equiv q^2/m_b^2$. 
The dashed line in (a) corresponds to the nonresonant invariant mass spectrum.
The Wolfenstein parameters are chosen to be $(\r, \eta) = (0.3, 0.34)$.
\item[\bf Figure 2] Branching ratio ${\mathrm{Br}}\,(B\to X_d\, e^+ e^-)$ (a) and \cp-violating asymmetry
$\asym$ (b) for $(\r, \eta) = (-0.07, 0.34)$. The dashed line in (a) represents the nonresonant spectrum.
\item[\bf Figure 3] Branching ratio ${\mathrm{Br}}\,(B\to X_d\, e^+ e^-)$ (a) and \cp-violating asymmetry
$\asym$ (b)  for $(\r, \eta) = (-0.3, 0.34)$.  The dashed line in (a) represents the nonresonant spectrum.
\end{enumerate}
%
%
%
%
%
%
%
%
\end{document}